**Title: The challenge and opportunities of quantum literacy for future education and transdisciplinary problem-solving**

**Authors: Nita, L., Mazzoli Smith, L., Chancellor, N. and Cramman, H.**

**Laurentiu Nita**
Department of Physics, Durham University, South Road, Durham, DH1 3LE
Email: laurentiu.d.nita@durham.ac.uk

**Dr Laura Mazzoli Smith (Corresponding Author)**
School of Education, Durham University, Leazes Road, Durham, DH1 1TA
Email: laura.d.mazzolismith@durham.ac.uk

**Dr Nicholas Chancellor**
Department of Physics, Durham University, South Road, Durham, DH1 3LE
Email: nicholas.chancellor@durham.ac.uk

**Dr Helen Cramman**
School of Education, Durham University, Leazes Road, Durham, DH1 1TA
Email: helen.cramman@durham.ac.uk

**Abstract**

Resulting from cross-disciplinary dialogue between physicists, computer scientists, educationalists, and industrial end users, we propose the concept of quantum literacy as one means of addressing the transdisciplinary nature of the complex problems that we see at the heart of issues around global sustainability. In this way, quantum literacy can contribute to UN Sustainable Development Goal 4, Quality Education. We argue that quantum literacy, as defined here, addresses the challenges of learning and skills acquisition within a highly bounded discipline and of access to the kind of powerful knowledge that should be more accessible to a wide group of learners throughout the life course, both students and professionals. Knowledge of quantum computing is arguably inaccessible to many, with knowledge of the complex mathematics involved a particular barrier to entry. Meanwhile it is increasingly important that the knowledge of quantum technologies is accessible to those who work with real world applications in an accessible way. We therefore argue for the importance of addressing pedagogic issues when powerful knowledge consists of dense concepts, as well as complex and hierarchical relations between concepts, in addition to presenting a strong barrier to entry in the form of mathematics. We introduce a specific puzzle visualization learning tool through which to achieve these pedagogic ends with respect to quantum computation. Visualization through puzzles can enable non-specialists to develop an intuitive, but still rigorous, understanding of universal quantum computation and provide a facility for non-specialists to discover increasingly complex and new quantum algorithms. Using the Hong-Ou-Mandel optical effect from quantum mechanics, we demonstrate how visual methods such as those made possible through the puzzle visualization tool, can be very useful for understanding underlying complex processes in quantum physics and beyond and therefore support the aims of quantum literacy.

**Main article body**

**Introduction**

> *The Sustainable Development Goals are the blueprint to achieve a better and more sustainable future for all. They address the global challenges we face, including those related to poverty, inequality, climate change, environmental degradation, peace and justice.*
>
> (https://www.un.org/sustainabledevelopment/sustainable-development-goals/)
>
> *In addition to improving quality of life, access to inclusive education can help equip locals with the tools required to develop innovative solutions to the world's greatest problems.*
>
> (https://www.un.org/sustainabledevelopment/education/)

In this position paper we present an educational concept and an associated learning tool that can contribute to United Nations Sustainable Development Goal (SDG) 4, Quality Education. We take this SDG as foundational, not only in terms of its inclusive aim, but also as being capable of addressing the access to powerful knowledge and understanding that will be needed at societal level in order to make greater progress in relation to other SDGs. We propose one innovative route to such improvements in knowledge and understanding, with an associated learning tool designed to circumnavigate the specialist knowledge that currently prevents wider understanding of, and hence applications for, quantum computation. We have developed this approach through cross-disciplinary dialogue between physicists, computer scientists, educationalists, industrial end users, and others and suggest that such cross-disciplinary dialogue is essential in the development of innovative ways of solving the most pressing societal problems, with quantum literacy being an example of this. Such cross-disciplinary dialogue is essential when considering the transdisciplinary nature of the complex problems that we see at the heart of issues around global sustainability. Indeed, in this position paper, we suggest that the highly bounded, specialist knowledge that characterises disciplines such as physics, need not prevent innovative educational approaches to widening the base of understanding in such fields, ensuring access for as many in society as possible. This position paper articulates the concept of quantum literacy as novel and important with respect to the UN SDG Quality Education. Also considered are the educational challenges presented, considering barriers to such learning and the description of a learning tool that can broaden access to this specialist field of knowledge.

**Quantum literacy**

Although quantum technologies are at an early stage, impacts they are having on our culture are already observable, with far-reaching benefits identified for a wide range of industries in the coming years, 'Quantum science promises to have a major impact on the finance, defence, aerospace, energy and telecommunications sectors'…'These technologies promise to change our lives profoundly' (QT SAB, 2015c: 4-5). The UK is part of a global race to industrialise quantum technology through the National Quantum Technology Programme (Knight and Walmsley, 2019). Industry managers and government will increasingly be expected to make decisions related to quantum computing technology and, we argue, they will have to become 'quantum literate' to avoid falling victim to misconceptions and hype. Yet at present, there is little understanding of, or expertise in, the skills required for effective quantum computational reasoning outside specialists in physics and mathematics. While quantum computing technologies provide an exciting new paradigm to approach some of the most difficult problems faced by humanity, how these technologies work is currently only understood by a small segment of the population, with strong perceived barriers to entry from outside of the field. It is recognised that investment in training and skills is needed;

> 'Both academia and industry need to play their part in developing a wide skills base not only in the physics but also in engineering, systems engineering, production engineering, business and entrepreneurship. Training should be provided both to young researchers and qualified people looking to adapt to new jobs and roles in quantum technology' (QT SAB, 2015c: 6).

This lack of understanding from outside of the field is problematic because many of the people who know the end use cases and the current state of the art classical computation techniques (application domain experts), which are currently used to solve them, will find it difficult to contribute. The only way quantum computing can succeed in the near term is to find the right niche areas to apply it and leverage the expertise of the people who currently solve these problems on how quantum technologies can make their methods better. Such a goal is fundamentally cross-disciplinary and cannot be achieved otherwise. Funding is largely directed towards highly specialist centres, for instance in the UK (QT SAB, 2015). There is also recognition of the need to 'invest in the development of a dynamic workforce that meets the needs of future industry' (QT SAB, 2015: 8) and the need 'to fund programmes that ensure a workforce with the right skills is available to companies as they seek to exploit these opportunities' (QT SAB, 2015; 22). Approaches therefore must include training, to some extent, far greater numbers of the current workforce to be quantum literate, as well as training a next generation of application domain experts to be quantum literate. We suggest that the concept of quantum literacy can help to structure the educational initiatives needed to support these approaches and also support the overarching aim through providing clarity of purpose. We advance a definition of quantum literacy as: *learning of the minimal body of fundamental knowledge of quantum mechanics that allows understanding of how quantum computation could be used in diverse application domains and the capability to assess claims related to quantum technology.*

Disciplinary-based approaches to dialogue across quantum computing and application domain experts can only go so far and we will need people with an understanding of both fields. This will in turn require innovative education and training tools and an increase in the level of general quantum literacy in the population. We argue that quantum literacy should aim to increase understanding rather than simply awareness. This understanding is important as quantum computing emerges as a technology and many non-experts are faced with decisions related to it, for instance a manager deciding if a quantum computation solution is right for a problem. This is the knowledge physicists and computer scientists do not have, i.e. knowledge not just about quantum computing, but the things quantum computing could be used for. Higher levels of quantum literacy will enable those from a wide variety of backgrounds to bring the potential benefits of dramatically faster and more complex data processing to their own businesses, industries and disciplines, enabling them to re-imagine the possibilities for data analysis and problem solving in their fields. The benefit of a broader range of people being able to access this understanding than is currently possible when quantum computing has to be learned through mathematics and physics, is therefore a wider understanding of the potential applications of quantum computing in diverse areas of society, which currently do not benefit from this technological revolution.

There are many concrete examples of areas in which quantum technologies could be game changing. The number of domains for which proof-of-concept quantum computing experiments have been conducted is far too many to list here, but includes subjects as diverse as computational chemistry (Kandala, 2017), flight gate assignments at airports (Stollenwerk et al., 2019), decoding of error correction codes (Chancellor et al., 2016), and hydrology (O'Malley, 2018). While these experiments are too early to show quantum advantage directly, there are areas in which provable (at least up to standard assumptions about computational complexity) quantum speedups are possible, for instance the famous Grover search (Grover, 1996), or Shor factoring (Shor, 1999) algorithms. Building on these insights and others, there have been many more algorithms with provable advantages developed, with applications in areas such as optimisation and cryptography. A review of such algorithms is provided by Montanaro (2016) and a running list of quantum algorithms by Jordan (2020). An equation free review focusing on continuous time quantum information processing is provided by Kendon (2020).

While quantum computing promises the broadest transformations of any quantum technologies, there are other quantum technologies which promise more near-term applications. These include the use of atomic systems to see through objects by detecting terahertz radiation (Downs, 2020), more accurate atomic clocks, which would allow the detection of small strains in the earth's crust for earthquake detection (Ludlow et al., 2015), and key distribution protocols for cryptography which can detect spying by construction (Minder et al., 2019). Since quantum mechanics equates to a different way of understanding reality, quantum literacy therefore potentially offers a different way of conceiving of problems in a range of diverse fields. In order to maximize the benefits of quantum computing and to reduce the risks, we need to up-skill and educate the current workforce and begin the training of our future programmers and decision makers in the counterintuitive ways of quantum computational thinking. Classical computer programs are not effective for quantum computing. Instead, quantum algorithms must be used which employ quantum phenomena such as interference of states. The problems for which quantum computation can offer solutions are themselves in need of constructing in light of the understanding that would be gained through quantum literacy.

**A social realist approach to scientific knowledge and understanding**

In this paper we focus on the educational challenge of quantum literacy and position the concept in social realist debates in the epistemology of knowledge. A social realist framing is useful in that it advances an argument which 'rehabilitates specialised knowledge and binds it back into a social framework on which it depends' (Young & Muller, 2013; 247), important given the transdisciplinary nature of the problems that quantum literacy could help to address. We too often lack a theory of knowledge (Maton, 2013), in terms of both the sociological structure of knowledge with properties, powers and tendencies, but also in terms of its intrinsic features. We argue that quantum literacy as we define it here addresses both the challenges of learning and knowing in a bounded discipline, as well as the relational structures of knowledge practices that create barriers to accessing areas of knowledge. We draw on an important debate in the field of education associated with the valuing of a particular systematic body of knowledge built over time, opposed, in current discourse, to seeing that knowledge in terms of whose interests it serves and how accessible it is, that is the argument about 'powerful knowledge' (Moore et al., 2006; Young, 2013; Young & Muller, 2013). As stated at the outset, we relate this to the UN SDG Quality Education in arguing the case for wider access to such powerful knowledge, which can be defined as:

> '(1) access to more reliable facts or truths; (2) access to higher level conceptual perspectives of the specialist field; (3) being able to see the specialist, structured form of a knowledge that differs from everyday experience; and (4) working with objective rather than learner-centred or social-interests-centred orientations to curriculum' (Yates & Millar, 2016).

If we are concerned to widen access to a body of systematic knowledge built over time, then we must also address the issue of disciplinary specialisation, with a different purpose and structure to non-specialised knowledge, requiring specialist institutions in which to develop and transmit it. A strong educational critique concerns elitism, in that, by definition, specialised knowledge will not be distributed equally and those who tend to have access to it are the already powerful (White, 2012). However, Young and Muller (2013) note the category mistake here, pointing out that knowledge of the powerful does not necessarily equate to powerful knowledge. Indeed, in our conceptualisation of quantum literacy, we would argue precisely that this powerful knowledge does not readily accrue to the knowledge of those with power to effect change. Young and Muller also answer the critique that education should be about the flourishing of society and the fostering of well-being, not knowledge acquisition *per se* (e.g. White, 2012) by saying that this poses a false dualism. We would concur and in the deliberate positioning of the concept of quantum literacy in this journal, directed towards the UN SDGs, we bring together the way in which access to and understanding of a specialised body of knowledge, in this case quantum computing, can be linked directly to societal well-being through support of the UN SDG Quality Education.

We advance the case of quantum literacy as axiomatic with respect to these and associated arguments, presenting both an area of knowledge, that of quantum computation, as powerful knowledge that should be more accessible to, and motivating for, a wider group of learners throughout the life course than is currently the case. We go on to suggest a specific learning tool through which to achieve these ends. The learning tool answers one further core issue that we think should pertain to all discussions of powerful knowledge if we have consensus about its role in curricula and that is understanding of the most effective pedagogies with which to teach the building blocks of knowledge so defined.

We assert that quantum mechanics - and quantum computation - represent a body of knowledge that cannot be thought of other than through the conceptualisation of powerful knowledge (Young & Muller, 2013; Wheelan, 2007) if we attend fully to the nature of this knowledge as specialised and its likely impact on human endeavour. Powerful knowledge takes account of how we differentiate knowledge and so is well placed to articulate the challenges posed by quantum literacy. Young and Muller talk about how we differentiate knowledge in many ways; epistemologically, aesthetically, morally. They give an apposite example, drawing on quantum theory, to elucidate how we differentiate knowledge from our opinions and experience in its recognition of a reality independent of us:

> 'Quantum theory is the most reliable theory of the physical world there has ever been and in that sense it is as near as we have got to physical reality. At the same time physicists do not know quite why it gives us such reliable predictions. Physics, like any powerful knowledge, pre-supposes that the natural world is real and that current knowledge is the nearest we get to what that reality is. At the same time, quantum theory is probably the knowledge most at odds with our everyday understanding: it tells us that the particular that constitute matter are in many places at the same time and that matter takes the form of a both a particle and a wave' (Young & Muller, 2013; 230).

Yates and Millar (2016) suggest that science, and physics in particular, would seem to be paradigmatic examples of powerful knowledge because of their strong disciplinary boundedness and vertical knowledge structure, in Bernstein's (1999) terminology. A key point that Young means to get across is the specialized and differentiated nature of this knowledge when compared with everyday experience and which we pick up on below in discussing a puzzle visualization tool designed to circumvent the challenge in learning associated with knowledge that is ostensibly counter-intuitive. Physics is a discipline described as being strongly hierarchical (Lindstrøm, 2000), utilising a vertical discourse that: '…takes the form of a coherent, explicit, and systematically principled structure, hierarchically organised, as in the sciences' (Bernstein, 1999: 159). Bernstein developed the idea of a hierarchical knowledge structure, which characterizes the natural sciences, to refer to how different knowledge structures build cumulatively and progressively, newer knowledge subsuming earlier knowledge and differing bodies of knowledge then differing in their degrees of verticality. The powerful knowledge we are focused on in relation to quantum literacy can be understood in this way, the verticality pointing to abstraction of real-world knowledge to decontextualized principles, often utilising dense nominalisations where one word comes to stand for a complex concept (Conana et al., 2016).

The challenge for education is therefore considerable and we demonstrate this idea of dense nominalisation by considering the concept of entanglement. Entanglement emerged through a thought experiment (Einstein et al., 1935) on the mathematics behind a predicted phenomenon in nature by the formulae of quantum physics. This thought experiment predicted a phenomenon so counterintuitive, that the paper presenting the thought experiment concludes, 'We are thus forced to conclude that the quantum-mechanical description of physical reality given by wave functions is not complete' (1935: 1). Entanglement as a word was later coined by Schrödinger, as 'the probability relations [when the quantum state of two corelated physical systems][…] is known by a representative in common' (1935: 1), and later observed in the laboratory to be correct. Einstein (1947 in 2001) referred to this as, 'spooky action at a distance'. The difficulty of explaining the concept of entanglement to others lies in the lack of analogies in human experience, with the concept itself being the result of applied mathematics and the coined word being a translation from the German word *verschränkung.*

In addition to the challenge of learning such concepts, physics is characterized as having a very strong hierarchical knowledge structure, so;

> '…difficulty in learning the subject does not lie in simply the number of concepts that need to be learnt, rather it lies in learning the myriad of relations among concepts… A hierarchical knowledge structure represents a way of knowing that is characteristic of physics…However, this structure is rarely (if ever) explicitly taught in physics' (Lindstrøm, 2000).

Morrow (1993) suggests that 'epistemological access' to the discourse is important, with discourse referring to how a discipline presents itself in the many symbolic modes it employs in addition to language. Students need to develop their ability to shift between these. In physics in particular, it is also suggested that students do not have access to the qualitative representational aspects that expert physicists do, approaching quantitative problems without this kind of additional understanding (Rosengrant et al., 2007; Conana et al., 2016). Rather, they adopt a formula-related approach, grappling with the mathematics straight away, without any recourse to some representation of the physical concreteness of the issue.

> '…many of the representational aspects of Physics tend to be taken for granted in teaching: although problem-solving is demonstrated in lectures, often the modelling and qualitative representational aspects are glossed over, and what students see written down by the lecturer is merely the mathematical representation of the problem situation…' (Conana et al., 2016; 32).

In Conana et al.'s study with university physics students, qualitative representations used to understand physical processes rather than just mathematical representations supported students' problem-solving practices, which were more congruent with how expert physicists would work. If we approach quantum literacy as an educational challenge, we then consider how to facilitate learning when powerful knowledge consists of such dense concepts, as well as complex and hierarchical relations between concepts and presents a strong barrier to entry in the form of mathematics. We do this by focusing on pedagogic goals to manage these challenges through discussing the benefits of a puzzle visualization tool.

Our understanding of quantum literacy can also benefit from work on transdisciplinarity, being concerned with overcoming obstacles to a wider understanding of knowledge pertinent to human existence through boundary crossing (Akkerman & Bakker, 2011), and boundary blurring (Klein 2013). Transdisciplinarity as concerned with complexity, multidimensionality and problem-focused research (Klein 2013) is therefore both conceptually and practically a means of facilitating quantum literacy. The ability to see problems in a transdisciplinary way and apply quantum computation techniques to solving them depends on understanding that is situated more broadly than within a single, bounded discipline. Klein refers to the transgressive imperative of transdisciplinarity, which can challenge disciplinary conventions and hierarchies of expertise through more participatory modes of knowledge across sectors. Drawing on hybrid modes of inquiry, practice and learning, transdisciplinarity:

> '...brings new objects into view, places practices in new configurations, contextualizes and resituates theory and learning, and heightens awareness of hybridization by incorporating once excluded forms of knowledge, including the understandings of lay people' (Klein 2013, p197).

In the puzzle visualization tool that we present, boundary crossing is facilitated through the removal of mathematics as a point of entry and through the provision of a learning tool that engages lay people in not only learning, but also the potential creation of real quantum algorithms.

We would propose Nowotny's (1993) 'protoexpert' as a tangible outcome of a more quantum literate society, someone that is able to operate effectively in the kind of transdisciplinary space that we describe above. These scientific protoexperts would possess some knowledge of quantum computation by virtue of understanding in addition to knowledge, to differing degrees, but sufficient so that this understanding and knowledge could then be applied in different disciplinary contexts to address varied domain-based problems. The rise of the protoexpert in other areas of science is noted, but quantum computation has not, as yet, been accessible to any extent that there could be said to be protoexperts in this field:

> 'The process of the social distribution of scientific knowledge itself is based on a general rise in educational standards and, more generally, on the success of the cultural, as well as the technological, message of science to society. It has led to a rising number of knowledge experts and protoexperts distributed throughout society, men and women who possess scientific and technological knowledge of different kinds and degrees and know how to apply them in different contexts, thus contributing to the production of novel configurations of knowledge and knowledge claims' (Nowotny, 1993: 308).

Probably the best example of protoexperts in classical computing are programmers, who often do not have formal training in computer science, but rather know how to write efficient code based on experience and incomplete understanding of low-level processes and then utilise programming skills in a range of domain areas. There are specific instances of lay communities shaping scientific research, such as Epstein's (1996) analysis of how AIDS activists transformed biomedical research practices in the field of AIDS research. An example of a well-known individual protoexpert is that of a computer scientist without formal training in biochemistry who went on to propose using DNA as a computational system (Adleman, 1994). This later become a separate field of study in its own right, called DNA computing, that lead to the invention of 'molectronics', the use of the complexity of DNA molecules in living systems to encode computation on (Amos, 2002).

**Mathematics as a barrier to entry**

> '...the relations between specialised and non-specialised knowledge differ in different disciplines. The boundaries between the two are for all practical purposes unbridgeable in physics and in the chemical and, increasingly, in the biological sciences, not the least as a result of the lack of ambiguity of the mathematics they use and the abilities they have developed to express the relationships between their concepts in precise mathematical form' (Young & Muller, 2013: 244).

The argument about powerful knowledge, introduced above, along with the necessity for precise mathematical knowledge needed in quantum mechanics, is a concern for many in that;

> '…although the high-status subjects such as physics and high-level mathematics may be intellectually powerful as well as socially and instrumentally advantageous, the abstracted detachment required for them produced disengagement by those from poorer backgrounds and reproduced the socially differentiated patterns of success and failure' (Young & Muller, 2013: 300).

Young sees this as a political issue, but we suggest here, through our example of the quantum visualization tool described below, that this can also be conceived of, and dealt with, as a pedagogical issue. Indeed Young (2013) suggests that pedagogy is still under-developed as a specialist field of knowledge. We suggest that we could interrogate the nature of how such specialist knowledge is taught by focusing more directly on pedagogy where the medium of learning – here mathematics – is itself a significant barrier to entry, taking on board Young's (2013) claim that,

> '…although knowledge can be experienced as oppressive and alienating, this is not a property of knowledge itself. An appropriate pedagogy, which engages the commitment of the learner to a relationship to knowledge…can have the opposite consequences – it can free the learner to have new thoughts and even think the 'not yet thought'' (2013: 107).

Yates and Millar state that 'One long-standing problem in the physics curriculum is the constitutive role that mathematics plays in physics' (2016: 305). It is relevant to note here the perceived link between this and suggestions that there are common misconceptions of some of the foundational concepts precisely because of an over-reliance on mathematics. Yates and Millar state that 'A reduction in mathematics is seen to provide room for a closer and more detailed conceptual understanding of such areas' (2016: 305), whilst at the same time a high level of mathematics is seen as being necessary for cutting edge topics like quantum mechanics. University physicists who were interviewed by Yates and Millar expressed concern, 'that students who spent all their time mastering the mathematics would not have the sense of the field or the creativity and initiative needed to take it forward' (Yates & Millar, 2016: 306).

Access to a particularly difficult notation language in mathematics effectively functions as a powerful and exclusive form of knowledge that is only accessible to those who have achieved well in it. Mathematical Dirac notation equations relay quantum states and concepts precisely, yet this is beyond the reach of most people. We can class this as a valued form of knowledge, which enables access to other learning and one in which issues of inclusion are therefore pertinent (Young & Muller, 2010). However, as suggested above, we can also class this as a form of knowledge wherein multiple barriers to learning pertain, resting primarily on 'the gate-keeping function of achievement in school mathematics' (Straehler-Pohl & Geller, 2013: 314). Mathematics is also a form of strong classification, consisting of a highly specialized discourse, with its own specialized set of internal rules (Bernstein, 1996). If we consider quantum computation as strongly bounded, then its use and possible contribution to society in terms of addressing the UN SDGs is limited accordingly.

We suggest that circumnavigating conventional forms of mathematics in order to work with the representations of quantum matter directly is a promising way to proceed towards quantum literacy and that doing so can be positioned as a direct challenge to a particular configuration of hierarchical knowledge. Added to the difficulty of accessing the mathematics is also the fact that these concepts are very difficult to communicate verbally, through linguistic description, because they are counter-intuitive, challenging the underlying common core of knowledge that non-specialists would access, derived from classical Newtonian physics. Here specialised knowledge also conflicts with lay, or common-sense knowledge. The issue therefore is not only one of knowledge acquisition, if knowledge is to be defined as knowledge-that (Ryle, 1946) but also understanding, if by understanding we mean something holistic, incorporating a creative act that links together knowledge of parts, imposes order, compares and contrasts (Cooper, 1995), incorporating knowledge-how. Ryle's distinction of knowing-that (propositional knowledge) and knowing-how (practical knowledge) is useful as quantum literacy clearly must incorporate both. We therefore also have to take on the challenge of how such understanding would have to sit alongside, and be compared with, every day or classical understandings of the physical world. This epistemology of understanding is therefore critical to the concept of quantum literacy. Cooper's definition of understanding is useful here and pertinent to quantum literacy and its dependence on both theoretical and applied knowledge:

> '…to compare and contrast, to amplify, abridge and paraphrase, to generalize and to instantiate, to emphasize, and so on, are all capacities which fall under understanding. Anybody who has at least some of these capacities is able not only to answer questions about the subject-matter of inquiry but also to raise new questions and so enlarge understanding' (1995: 209).

**Quantum puzzle visualisation tool**

Learning through some activity, for instance playing a game in a socially scaffolded environment, draws on Vygotsky's social-constructivist theory of learning (1997) where social interaction plays a fundamental role in the development of cognition. By engaging in investigative puzzle-solving activities, students acquire new understanding by actively constructing their own knowledge through experience. Using puzzles relies on problem-solving as a means through which concepts that are otherwise challenging, can be learned. 'Game-based learning' uses gamified content to meet instructional goals (Zainuddin et al., 2020) with three positive themes relating to the use of gamification in education evidenced in the literature: learning achievement, motivation and engagement, and interaction and social connection. When designing games for the education context, it is essential to consider the learning or behavioural outcomes of the gamified task (Schöbel, 2020). Whether learning outcomes have been achieved or enhanced through the use of the gamified element must be carefully evaluated against the intended learning or behavioural outcomes. Studies on teaching various aspects of quantum mechanics have shown that learning through gamified means can boost motivation for learning and improve learning outcomes (Eggers Bjælde et al., 2015). It is also shown that such methods can foster cross disciplinary collaboration (Magnussen, 2012).

Making use of puzzle games to describe physical phenomena has been shown to be very good at explaining certain observations in physics, such as remote optimization of ultracold atoms in an experiment by experts and citizen scientists (Heck et al., 2018), quantum speed limit (Sørensen et al., 2016), and quantum simulations (Lieberoth et al., 2015). These games are specific to solving certain problems however, while the puzzle tool discussed in this paper differs in scope, in being complete, to integrate and describe visually the body of quantum physics. It therefore stands as an alternative method to create any types of puzzles based on quantum physics, while solving the puzzles allows the player to observe the dynamics and learn the methods to resolve the puzzle without having to also understand the mathematical framework behind it. Scaffolding towards the mathematics behind the puzzles can be done through using the puzzle tool, yet it is not necessary in order to find a solution to a puzzle, since both methods are complementary, as described below.

The puzzle visualization tool we introduce is not the first work to express the mathematics of quantum mechanics in a visual way. For example, there is significant work being done on graphical calculi such as the ZX graphical calculus (Coecke & Duncan, 2011) and related graphical calculi (Backens & Kissenger, 2019). In fact, matrix operations have even been represented in a similar fashion to our tool within the graphical calculi community (Zanazi, 2015; Sobocinski, 2015; Bonchi et al., 2017).  Like our tool, these graphical calculi are mathematically rigorous, however, their goal is entirely different. While the goal of our techniques is to educate quantum non-experts on quantum computing, the primary goal of the graphical calculus community is to provide more powerful tools to those who are already mathematical experts. One exception to this pattern is the graphical tool developed in Roffe et al. (2020), which was developed not for all quantum mechanics, but for the design of quantum error correction codes.

We suggest that the entire body of knowledge needed to learn and work with quantum computation can be acquired through a blend of intuition derived from the process of engaging with puzzle playing, alongside scaffolded transmission of conceptual knowledge at particular stages of playing. We take a multifaceted conceptualisation of what and how learning may unfold, drawing on understandings of the importance of active learning in science (Wieman & Perkins, 2005; van Heuvelen, 1991). The ultimate aim of creating this new learning approach for quantum computation, pioneered by *Quarks Interactive*[1], is to enable non-specialists to develop an intuitive, but still rigorous, understanding of universal quantum computing and to provide a facility for non-specialists to discover increasingly complex and new quantum algorithms.

**The workings of the quantum puzzle visualisation tool**

In this quantum puzzle visualization tool, the visuals are not an approximation of reality, but represent what actually happens in the quantum world. The game engine is based on the matrix mathematics which sits behind quantum state representations, but the mathematics takes a visual, rather than equational, form. This representation is also complete in the sense that it can represent all isolated quantum states. As stated above, quantum computing is difficult to understand because the underlying principles and concepts deviate from the laws of nature and logic that we usually defer to, unlike in classical computing. Many important problems in quantum computing are uniquely related to quantum physics and teaching classical computing first could be confusing because of the extent of the difference. For example, classical probabilities can only add, the possibility of reaching an outcome in an additional way can only increase probability of that outcome happening; counter-intuitively quantum probability amplitudes can cancel each other, and an additional way to reach an outcome can lead to destructive interference and cause the probability to decrease, possibly even to zero. It is these new principles which can be very difficult to describe in words and traditional equations. A visualization tool circumvents both these issues, enabling learners to gain an intuitive grasp of concepts such as superposition and phase interference visually, as visualization enables a more immediate grasp of complex states and concepts which challenge everyday understanding. The abstraction used can convey the entire range of quantum phenomena, up to numerical precision over the dynamics of small universal quantum computation systems.

The visuals used in the puzzle tool are based on an entirely graphical version of the matrix-vector representation of the mathematical state space description of full systems used by quantum physicists to design quantum algorithms. Quantum physicists use matrix-vector multiplication in order to calculate the effect of a change (described by a matrix of complex numbers) on a quantum state (described by a vector of complex numbers). Multiple such changes executed in a specific order on a quantum state is what physicists do in order to create the actual quantum circuits used in computation. Effectively the matrices used in universal quantum computation calculus are represented as edges on a bipartite graph (a visual map). The colours and the sizes of the balls used in the puzzle tool to represent the quantum states are the exact encoding of the complex numbers used to define such quantum states, with colour representing phases, and size representing amplitude of the complex numbers. Hence the graphs used in the puzzle tool and the actual matrices used in performing quantum computation contain the same information, with the only difference being the way they are represented: the first being purely visual and intuitive to use (as any visual puzzle would be), whilst the second requires knowledge in mathematics to understand, process and make use of it.

---

[1] *Quarks Interactive* has developed an innovative puzzle visualization tool to explore the process by which non-specialists in quantum mechanics develop their understanding of quantum computational thinking.

This aspect is furthermore defined in the puzzle tool as an interaction between the balls representing quantum states and the graphs in the tool representing the matrices, which is completely equivalent to the matrix-vector multiplication done by physicists. Because the translation of matrices to visual elements is exact, this representation is also exact. The order of such applied changes to the quantum states (the order of the matrix-vector multiplications) as the physicist is designing a quantum circuit is represented in the puzzle tool in real time through the evolution of the balls and graphs. The balls always pass through the graphs from the beginning to the end of the graph (which is identical information to the series of calculations a physicist would do, but displayed visually), representing the order and the dynamic of the quantum circuit as it is being created. The player is allowed to perform changes to the graphs by adding puzzle pieces (matrices in visual form representing state changes) at any point in the circuit, equivalent to a quantum physicist adding changes in a different order to quantum states with the aim of designing a quantum circuit to achieve a goal. In the puzzle tool, the goal itself is described as the number, position, colour and size of the balls that should arrive at the end of graphs. Using this tool, problems known by physicists as state compilation or decomposition problems can be performed entirely visually, without the need for prior knowledge in the field, allowing non-experts to create their own quantum algorithms once they are familiar with the visual representation.

We propose that from this visualization tool the players will be able to learn about fundamental principles behind quantum mechanics such as superposition and interference, through using trial and error as they attempt to solve puzzles that make use of such phenomena in physics. Entanglement and superposition are phenomenon occurring in nature that are at the core of the theoretical framework within quantum mechanics, hence because the puzzle tool is representing this framework, such phenomenon are also present as visual outcomes of the dynamics that the player is able to set up, and make use of, for creating quantum circuits. Because high level tools for quantum computing are not yet fully developed, understanding the underlying building blocks is crucial, unlike in classical computing where much of the low level behaviour of the computer can be abstracted away.

The dynamics of classically counter-intuitive processes such as phase amplification can be understood in such a way that even if the mathematics is hard to grasp, they can be intuitively grasped by engaging with the visual tool and solving puzzles. The visualization tool is specifically of something real i.e. quantum circuits, which include non-Clifford[2] gates and are therefore universal for quantum computing. The fact that the game is a full, exact representation of quantum mechanics, necessarily limits the systems to small sizes (if the game itself could exactly simulate large quantum computers, we would not need large quantum computers), however these small sized examples can build intuition for larger systems, which could not be represented in the game. This is therefore a gateway to further learning because it presents complex numbers and linear algebra in a much more accessible, that is purely visual, method.

**The difficulty in understanding quantum mechanics processes in language and mathematics**

Since quantum mechanics is the body of knowledge that attempts to offer a full description of reality, it is of mass appeal, as discussed above, often attracting interest from those not versed in the level of mathematics and physics required to fully understand the information presented, and misconception can easily be generated. We will show why this is the case, by describing the Hong-Ou-Mandel optical effect (Hong et al., 1987) from quantum mechanics. We do this first by analogy with a classical example - throwing balls on a splitter – followed by a simplified mathematical description of the actual quantum experiment, which involves throwing photons (light particles) at a beam splitter[3]. We then show how the same body of information can be conveyed in a more efficient manner and with a lower barrier of entry using the puzzle visualisation tool.

---

[2] Clifford gates are a subset of quantum gates (complex matrices that impose a change to a quantum state), which are efficiently classically simulable; the inclusion of at least one non-Clifford gate makes the circuits hard to simulate classically. In the puzzle tool, these are the puzzle pieces the player can place in any order.

One of the most common misconceptions is that quantum computation works by simply trying all possibilities at once and taking the right solution. While this statement has a kernel of truth in it, the reality is not quite this simple. While quantum systems can exist in superpositions of different classical states, reading out a solution requires the use of a phenomena called interference. We know that classically, probabilities can only add, an additional route to an outcome makes that outcome more likely, and never less. Let's take as an example a situation in which we are dropping two bouncing balls right on the top of a perfectly positioned splitter that is hitting the centre of each ball. The balls can bounce either left or right after hitting the splitter. The drawing below depicts the physical process of dropping the two balls (the upper arrow) on a splinter (the purple symbol), with the lower arrows representing the direction of the bounce that follows. We expect the probability outcome of repeating the experiment multiple times of dropping the balls and measuring the direction of how the balls will bounce, to follow one of the three scenarios:

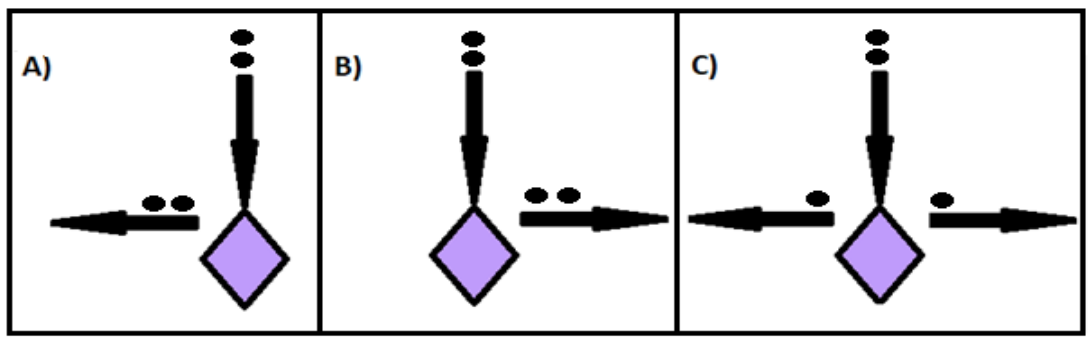

*Figure 1: Balls are represented with the black dots and arrows show direction of bounce*

A) 25% chance for both balls to bounce left.
B) 25% chance for both balls to bounce right.
C) 50% chance that each ball will bounce in an opposite direction.

If the balls behave quantum mechanically and are indistinguishable photons, upon measuring, the result will no longer follow the example above. This is known as the Hong–Ou–Mandel optical effect and this cannot be understood without quantum interference, for which there is no classical analogue. This is also one of the simplest quantum interference examples that can be observed in a laboratory, hence we will focus on this example to show how using visual cues can enable this to be understood more effectively. The outcome for what will happen to our quantum balls is a 50% probability of either of the following two events happening:

A) Both balls go left.

---

[3] A beam splitter is the device that forces the photon to either bounce off or pass through it, by analogy going either left and right with the classical example.

B) Both balls go right.

*The Hong–Ou–Mandel effect explained mathematically*

This analogy to the classical world is necessary because in quantum mechanics we are dealing with quantum objects that defy our intuition of how they might behave. Our understanding of quantum mechanics shows that probabilities can also decrease, or cancel even to reach zero, because of the effects of quantum interference, a phenomenon required to explain the Hong–Ou–Mandel optical effect. The understanding of how the world behaves is counter intuitive at tiny scales.

To understand this effect, first we must understand the behaviour of a single photon. We send a single photon through the beam splitter (Eq. 1) that has a 50% chance to allow the photon to pass through. We use the quantum mechanics formalism called Bra-Ket notation to encode our events and their probability for it to not pass through as $|0\rangle$ and for it to pass as $|1\rangle$.

$$H_{bsp}= = \frac{1}{\sqrt{2}}\begin{pmatrix} 1 & 1 \\ 1 & -1 \end{pmatrix}$$

(Eq. 1)

In this case the $|0\rangle$ state corresponds to the photon travelling vertically, and the $|1\rangle$ state corresponds to it travelling horizontally. Note that the factor of -1 is necessary for the beam splitter to be unitary. Our starting state before we send the photon, in Bra-Ket $|\psi_{in}\rangle$ is a sum between the events that it did not pass through with 100% probability and that it did with 0% probability, as seen in Eq. 2.

$$|\psi_{in}\rangle = 1|0\rangle + 0|1\rangle$$

(Eq. 2)

Now multiplying the matrix from Eq.1 with Eq.2 we can get the probability distribution for the photon to be in both states, $|\psi_{out}\rangle$, which normalized gives us the 50% split of events in Eq. 3. A beam splitter that performs such an operation is called a Hadamard operator in quantum computation and its effect is to place a quantum particle in a superposition state, pre-measurement.

$$H|\psi_{in}\rangle = |\psi_{out}\rangle = \frac{1}{\sqrt{2}}(|0\rangle + |1\rangle)$$

(Eq. 3)

The Hong–Ou–Mandel effect occurs in the case when we have two identical photons that are going through a beam splitter at the same time. For this quantum effect to happen, we must have identical photons. The condition for a pair of identical photons (recall that photons are Bosons) is that their collective wavefunction is an eigenstate of the swap operator with a +1 eigenvalue[4],

$U_{swap}|\psi_{init}\rangle=|\psi_{init}\rangle$, where $U_{swap}= \frac{1}{\sqrt{2}}\begin{pmatrix} 1 & 0 & 0 & 0 \\ 0 & 0 & 1 & 0 \\ 0 & 1 & 0 & 0 \\ 0 & 0 & 0 & 1 \end{pmatrix}$ and our initial state is described by Eq.4 using the same Bra-ket method as for a single photon, with the difference that now we are representing the behaviour of two photons in the same formula. If the description of the initial state of the photon before passing through the experiment is Eq. 2, for two photons we define the initial state in Eq. 4. The states that |01⟩, |10⟩ form the initial state, with one photon travelling vertically and one travelling horizontally. The state |00⟩ represents the case where both photons are travelling vertically and |11⟩ represents both travelling horizontally[5]. We however consider an initial condition where the photons are travelling perpendicular to each other. Our quantum state now is the sum of two ways of labelling the photons, labelling the horizontally (vertically) travelling one as the first photon and therefore the vertically (horizontally) travelling one as the second photon, shown in Eq.4. Note that the relative phase between the photons must be positive so that the eigenvalue with respect to swapping is +1.

---

[4] This is the physics knowledge required *a priori* to understand the experiment, set out here to make our point.

$|\psi_{init}\rangle = \frac{1}{\sqrt{2}}(0|00\rangle + 1|01\rangle + 1|10\rangle + 0|11\rangle) = \frac{1}{\sqrt{2}}(|10\rangle + |01\rangle)$

(Eq. 4)

In the case of these two identical photons, the effect of our quantum beam splitter is described by the tensor product $U_{bsp}$ in Eq. 5 of two Hadamard operators shown in Eq. 1.

$U_{bsp} = H_{bsp} \otimes H_{bsp}$

(Eq. 5)

When the beam splitter comes into effect, our $|\psi_{out}\rangle$ defines the outcome as in Eq. 3, $\psi_{out}\rangle = U_{bsp}|\psi_{init}\rangle$. Our final state is again the sum of all probabilities of all events happening. Here is where we can see the effects of quantum interference effects mathematically, because some of the probabilities of the events have negative signs in Eq. 6, and the result shown in Eq. 7 gives us the outcome of the experiment in mathematical form.

$|\psi_{out}\rangle = \frac{1}{2\sqrt{2}}|00\rangle + |00\rangle - |01\rangle + |01\rangle + |10\rangle - |10\rangle - |11\rangle - |11\rangle)$

(Eq. 6)

$|\psi_{out}\rangle = \frac{1}{\sqrt{2}}(|00\rangle - |11\rangle)$

(Eq. 7)

We get a phase with each reflection that leads to quantum interference between probability amplitudes of different events in particular, the two different ways for one photon to travel in each direction after the output cancels, and the photons therefore must either both be travelling vertically or both be travelling horizontally after the beam splitter. This core concept of interference explained in mathematics and witnessed in nature, shows how counter intuitive quantum mechanics can be. We propose that it might be much simpler to explain this effect visually, removing the need for an understanding of mathematical Dirac formalism which is an entry barrier, and which we suggest makes quantum computation harder to understand than in a graphical representation.

*The Hong–Ou–Mandel effect explained through the puzzle visualisation tool*

Here we discard the mathematical Dirac formalism and encode this experiment in the quantum puzzle visualization tool.

---

[5] An important aspect of working with quantum computation is to be able to encode physical phenomena (or world problems of any kind) in quantum states represented by bitstrings of 0`s and 1`s, to which other bitstrings of 0`s and 1`s are the solution to the problem. The quantum algorithm is the process of finding that solution.

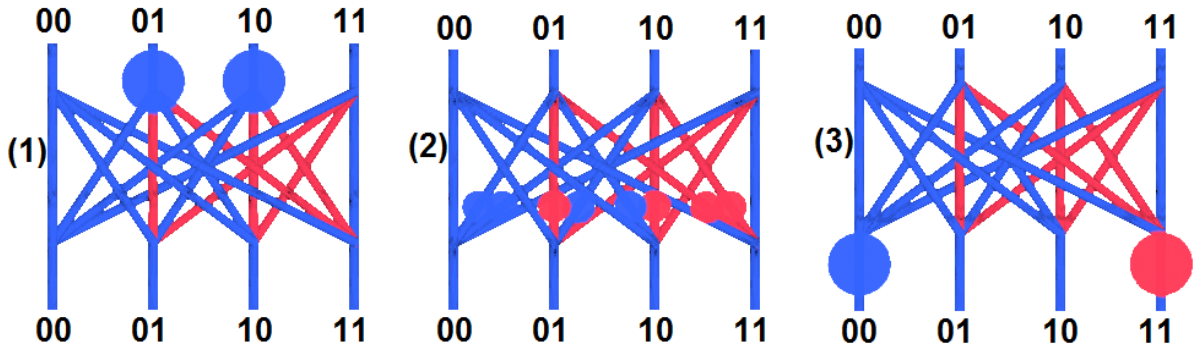

*Figure 2: The Hong-Ou-Mandel experiment realised in the puzzle visualization tool*

Figure 2, series (1) to (3), show a series of snapshots from a dynamic, visual representation of the same topic as described in the previous section. Our two photons are represented as two blue balls (phase is encoded in colour, these photons are identical hence their phase is identical), defining the $|\psi_{init}\rangle$ quantum state from Eq. 4, shown visually in Figure (1). The position of the photons is encoded in bitstring format (the same principle as with the Dirac notation from the previous chapter). As before, we consider two photons, hence our quantum state is defined by the position of both, using the falling 'quantum balls' from the puzzle tool.

In the snapshot in Figure (1), we chose to represent a photon falling vertically with bitstring 01, starting to traverse the graph, whilst a photon traveling horizontally starts at bitstring 10, both identical in size and colour. By using the bitstring encoding and representing the change effect as a graph, we can represent all possible combinations of photons as they travel as a graph, starting with 00 (if both would enter the beam splitter vertically), and ending with 11 (if both would enter horizontally). Because the representation in the puzzle tool is dynamic, three different static snapshots from the puzzle tool are needed to fully represent the dynamic visually of the mathematics behind the beam splitter. The bitstrings at the top, in Figure (1), define their starting positions: in our case, 01 and 10, because the photons are fired from opposite directions at the start, therefore we do not expect both photons to arrive simultaneously at the splitter from either vertically (00) or horizontally (11).

Figure (2) shows the dynamics of the change effect (the matrix-vector multiplication) performed by the beam splitter on the quantum state defining the position of the photons throughout the entire evolution process. With this method the player can see the actual effect in a visual instead of calculus. Each of the blue balls split to smaller, equal sizes. The horizontal photon (from 01) splits and becomes two blue balls travelling towards bitstrings 00 and 10 and two red balls towards 01 and 11, whilst the horizontal photon (from 10 splits) in two blue balls towards 00 and 10 and two red balls towards 10 and 11. This is the exact visualisation of the change effect mathematically described by Eq. 6, with the difference being that the complex numbers defining the matrix-vector multiplication are entirely achieved through visual indicators instead of mathematical notations.

The final quantum state, $|\psi_{out}\rangle$ is depicted in Figure (3) and it represents the outcome put forward in Eq 7, after the quantum states of different phases annihilate/ amplify, as described in Eq. 6. The phase that leads to the cancellation of quantum states within the effect of the beam splitter can be thus observed. Quantum interference can be seen in the cancellation of blue and red balls in Figure (2). On the other hand, the pair of blue balls and pairs of red balls are reinforced within bitstrings 00 and 11. We remember that bitstring 00 is equivalent to saying both photons travel vertically while 11 is for horizontally, hence the visual gives us a full representation of the behaviour before the beam splitter (Figure 1), the quantum effect of the beam splitter (Figure 2) and of the outcome, after the beam splitter (Figure 3), in accordance with the physical experiment.

We suggest this as a more accessible method to create intuition and understanding of the effect of the beam splitter proposed in the Hong-Ou-Mandel experiment. The puzzle tool allows the creation of any quantum circuit with the ability to witness in real time the effects of any change in its configuration. We selected this experiment for its simplicity in conveying why visual methods such as these can be very useful for understanding underlying complex processes in quantum physics and beyond.

**Concluding comments**

Development of quantum computing tools has mostly been restricted to research laboratories and universities over the past decades. This effort has given us an excellent understanding of how the technology can be used in a decade or so, once the hardware has fully matured. It has also given us prototype hardware available now, on the verge of being able to do things that cannot be reproduced classically. This leaves us with a gap, known as the Noisy Intermediate-Scale Quantum (NISQ) era (Preskill, 2018), in which we have hardware to use but need to develop new and tailored techniques to use it. Experts in quantum computing are best poised to find ways to exploit this era, but there is also the possibility that the 'killer app' for the NISQ era will come from a newcomer to the field. To encourage this, quantum hardware manufacturers such as IBM are currently working to make the hardware available for people to use and companies such as *Quarks Interactive* are providing alternative software methods to help them do it. An important factor in how effective newcomers can be is the quality of the educational materials that will guide them.

As discussed above, the bounded knowledge of quantum computing in relation to knowledge of real world applications held by those who solve domain-based problems in practice, and from those making the higher level decisions about quantum computing, will lead to inevitable inefficiencies and the potential for a large number of 'false starts' and other misunderstandings. Equally, managers who are put in the position of making decisions about research, development, and deployment of quantum computing, are likely to make suboptimal decisions and in particular, it then becomes very difficult for them to set realistic expectations for progress and potential applications. This leads to the potential of a 'quantum winter', where misunderstanding of the promises from quantum computing leads to a downturn in investment in the field, thereby delaying the realisation of large-scale useful quantum computing. We propose quantum literacy as a useful concept through which to take account of the normative theory of expertise in these fields, foregrounding pedagogic issues and the epistemological features that pertain and in so doing, facilitate access to the powerful knowledge that underpins quantum technologies.

**Competing Interests**

Laurentiu Nita is the founder of Quarks Interactive.

**Grant information**

All authors were funded by UKRI (United Kingdom Research and Innovation) grant number BB/T018666/1. Additionally, NC was supported by UK Engineering and Physical Sciences Research Council Grant number EP/S00114X/1 and LN was supported by a Durham University studentship.

**Acknowledgements**

The authors thank James Wootton for many useful discussions and for critical readings of the paper.